\documentclass[aps,prb,twocolumn,superscriptaddress,showpacs]{revtex4-1}
\usepackage{graphicx}
\usepackage{dcolumn}
\usepackage{bm}
\usepackage[usenames,dvipsnames]{color}
\usepackage{amsmath}
\usepackage{amssymb}
\usepackage{amsfonts}
\usepackage{mathrsfs}
\usepackage{epsfig}

\begin{document}

\title{Gate-tuned Differentiation of Surface-conducting States in
Bi$_{1.5}$Sb$_{0.5}$Te$_{1.7}$Se$_{1.3}$ Topological-insulator Thin Crystals}

\author{Janghee Lee}
\author{Joonbum Park}
\author{Jae-Hyeong Lee}
\author{Jun Sung Kim}
\author{Hu-Jong Lee}
\thanks{hjlee@postech.ac.kr}
\affiliation{Department of Physics, Pohang University of Science and
Technology, Pohang 790-784, Republic of Korea}

\date{\today}

\begin{abstract}
Using field-angle, temperature, and back-gate-voltage dependence of
the weak anti-localization (WAL) and universal conductance
fluctuations of thin Bi$_{1.5}$Sb$_{0.5}$Te$_{1.7}$Se$_{1.3}$
topological-insulator single crystals, in combination with
gate-tuned Hall resistivity measurements, we reliably separated the
surface conduction of the topological nature from both the bulk
conduction and topologically trivial surface conduction. We
minimized the bulk conduction in the crystals and back-gate tuned
the Fermi level to the topological bottom-surface band while keeping
the top surface insensitive to back-gating with the optimal crystal
thickness of $\sim$100 nm. We argue that the WAL effect occurring by
the coherent diffusive motion of carriers in relatively low magnetic
fields is more essential than other transport tools such as the
Shubnikov-de Hass oscillations for confirming the conduction by the
topologically protected surface state. Our approach provides a
highly coherent picture of the surface transport properties of TIs
and a reliable means of investigating the fundamental topological
nature of surface conduction and possible quantum-device
applications related to momentum-locked spin polarization in surface
states.

\end{abstract}


\date{\today}
\maketitle
\section{Introduction}
Similar to an ordinary-band insulator, a topological insulator (TI)
has a bulk energy gap in its band structure, which is generated by a
strong spin-orbit interaction. The topological phase transition,
brought about by the band inversion in the material, induces
Dirac-fermionic surface-conducting channels.
\cite{Fu045302,Fu106803,Zhang438,HasanRMP3045,QiRMP} This
topologically protected surface state (TSS) has a helical spin
texture that is robust to small perturbations conserving the
time-reversal symmetry, and thus prohibiting backscattering by
nonmagnetic impurities.
\cite{RoushanNature1106,ZhangPRL266803,AlpichshevPRL016401}

Diverse transport studies were conducted to characterize the TSS. In
general, however, as-grown TIs are $n$- or $p$-doped so that the
surface conduction can be predominated by bulk conduction.
\cite{XiaNatPhys,ChenScience,HsiehNature} Efforts have been made to
reduce the bulk conduction by tuning the Fermi level ($E_F$) into
the bulk band gap.
\cite{ChenScience,HsiehNature,RenPRB,TaskinPRL,Arakane,RenPRB155301,
ChenPRL176602,CheckelskyPRL196801,KimNatPhys460,SteinbergNano5032,KongNatNano705,
HongNatComm757,WangNanoLett1170} Even with these efforts, however,
critical inconsistencies were present in the previous transport
measurements. For instance, in bulk TIs with a thickness larger than
$\sim \mu$m, two-dimensional (2D) Shubnikov-de Haas oscillations
(SdHO) were observed. Nonetheless, the weak anti-localization (WAL)
effect, relevant to the TSS, was often absent in the corresponding
measurements, or, if present, did not fit well to the 2D
Hikami-Larkin-Nagaoka (HLN) WAL expression
\cite{AnalytisPRB205407,RenPRB195309,AnalytisNatPhys960,RenPRB241306,PetrushevskyPRB045131,XiongPhisicaE917,QuScience821,
RenPRB075316,RenPRB155301,CheckelskyPRL246601,HLN} (see Appendix \ref{A}).
In a TI, the WAL effect is generated by a
strong spin-orbit interaction and the consequent destructive
interference between two electron waves traveling along a diffusive
closed path in a time-reversal manner.
\cite{AndersonPR1492,HLN,Bergmann}

These inconsistencies between the 2D SdHO and WAL were also observed
in thinner flakes, with a thickness less than $\sim \mu$m.
\cite{HongNatComm757,WangNanoLett1170,XiuNatNano216,HeNanoLett1486}
Furthermore, previous 2D-SdHO observations
\cite{RenPRB155301,WangNanoLett1170,AnalytisNatPhys960,RenPRB241306,PetrushevskyPRB045131,
XiongPhisicaE917,XiuNatNano216,HeNanoLett1486} may not have been
fully relevant to the surface conduction by the TSS.
\cite{PetrushevskyPRB045131,CaoPRL216803,XiongPRB045314} Accurately
identifying the Berry-phase shift associated with the TSS requires
measurements in very strong magnetic fields, with careful
Landau-level indexing. \cite{XiongPRB045314} In most of the previous
studies, however, the $1/2$ Berry-phase shift was determined based
on observations in relatively weak magnetic fields.
\cite{RenPRB155301,WangNanoLett1170,RenPRB241306,XiongPhisicaE917,XiuNatNano216,HeNanoLett1486}
Ambipolar characters with back-gating were also observed in the
transport of TIs, which were assumed to be associated with the TSS.
Here, however, the WAL effect was absent in the samples with
relatively high carrier densities.
\cite{SteinbergNano5032,KongNatNano705,HongNatComm757} The WAL
effect observed in some of these ambipolar-transport samples were
reported to arise from the coupling between the surface and the bulk
bands, rather than the TSS exclusively.
\cite{ChenPRL176602,CheckelskyPRL196801,KimNatPhys460,ChenPRB241304,
KimPRB073109,MatsuoPRB075440,SteinbergPRB233101,ZhangAFM2351,ChaNanoLett1107}

It is an extremely difficult task to reliably separate the TSS from
other conductance contributions. In this study, we minimized the
bulk conduction using high-quality
Bi$_{1.5}$Sb$_{0.5}$Te$_{1.7}$Se$_{1.3}$ (BSTS) TI single crystals,
with $E_F$ lying in the bulk gap without gating. We confirmed that
the WAL effect and universal conductance fluctuations (UCF) indeed
arose from the top and bottom surfaces. By back-gate tuning the WAL
characteristics, we identified the TSS conducting characteristics
and the coupling between the TSS and the topologically trivial
two-dimensional electron gas (2DEG) states that emerged due to band
bending near the bottom surface. The ambipolar Hall resistivity of
the bottom surface was consistent with the back-gate-voltage
($V_{bg}$) dependence of the longitudinal resistance of the TSS.
This study provides a reliable means of differentiating the TSS of
TIs from those of the bulk conducting state and the topologically
trivial 2DEG states, along with a highly coherent picture of the
topological surface transport properties of TIs.

\section{Sample preparation and measurements}
BSTS single crystals were grown using the self-flux method.
\cite{TaskinPRL,RenPRB} Stoichiometric mixture of high-purity
starting materials (Bi(5N), Sb(5N), Te(5N), Se(5N)) were loaded in
an evacuated quartz ampoule, which was then heated up to 850
$^\circ$C. After annealing at 850 $^\circ$C for 2 days to enhance
the material homogeneity, the melt mixture was slowly cooled down to
600 $^\circ$C for a week. Before complete furnace cooling it was
kept at 600 $^\circ$C for one more week to further improve the
crystallinity. The stoichiometry and the high crystallinity of the
single crystals were confirmed by the energy dispersive spectroscopy
and the x-ray diffraction, respectively.

The bulk transport properties were examined using $\sim$100
$\mu$m-thick cleaved bulk crystals. For detailed characterization of
transport properties with back-gating, BSTS flakes, which are 22 to
230 nm in their thickness, were mechanically exfoliated onto a Si
substrate capped with a 300-nm-thick oxidized layer. This was then
followed by standard electron (e)-beam patterning and e-gun
evaporation of Ti/Au (10 nm/100$-$350 nm thick) bilayer electrodes
and contact leads. For thick crystals, the electrode contacts were
prepared using silver paste. In total, four thick bulk crystals and
six thin flakes were investigated using standard lock-in
measurements, varying $T$ from 290 to 4.2 K.

\section{Results and discussion}
\subsection{Thickness and temperature dependence of resistance}

\begin{figure}
\includegraphics[scale=0.41]{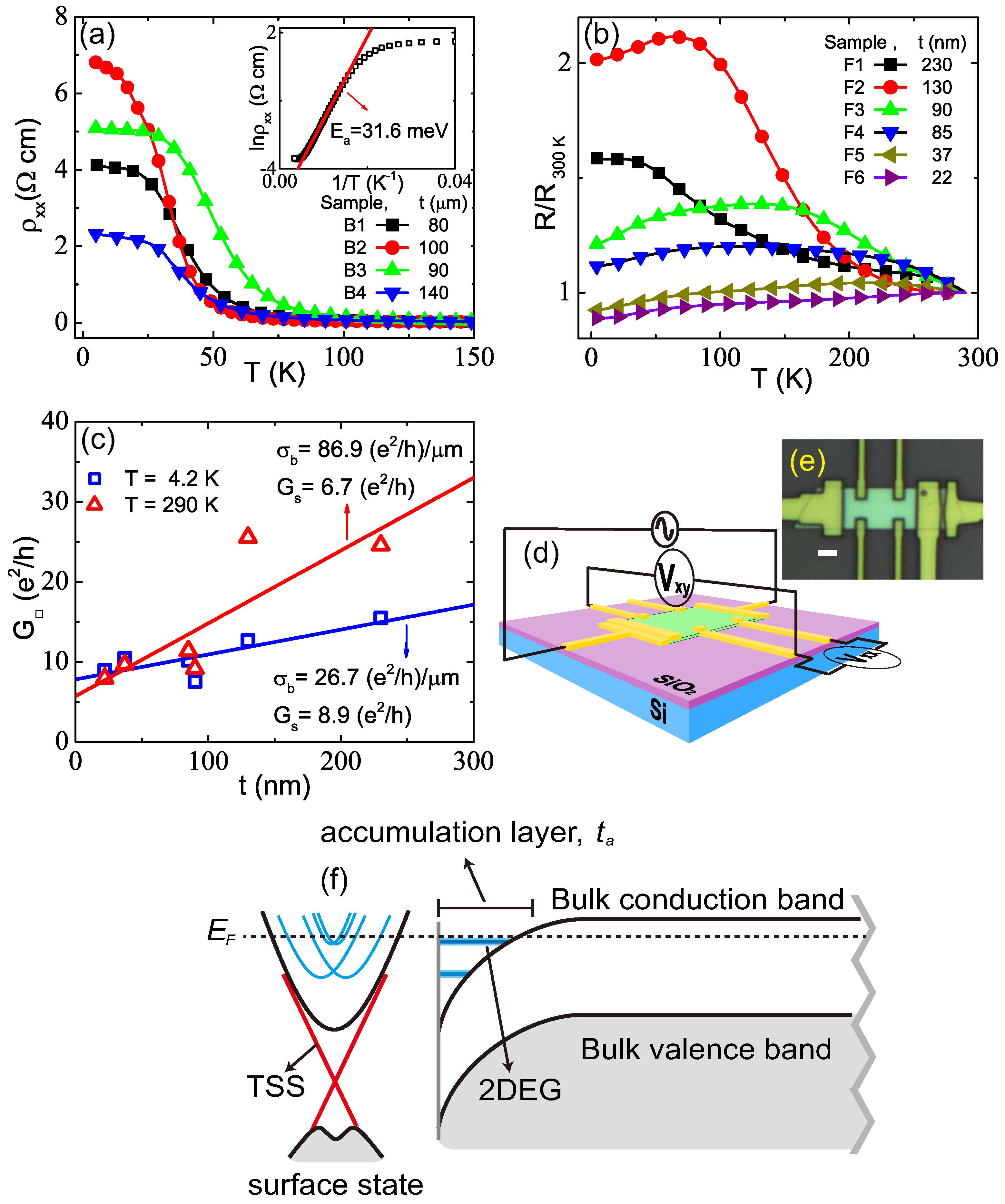}
\caption{(Color online) (a) $T$ dependence of the bulk crystal resistivity
of the TI. Inset: Arrhenius-law fitting for the sample B3.
(b) $T$ dependence of the normalized resistance of thin
flakes. (c) Thickness dependence of the sheet conductance
of thin flakes at 290 K (triangles, red online) and 4.2 K (squares, blue online).
Solid lines are best fits to Eq.\ (\ref{SheetConductance}). (d) Schematic
measurement  configuration and (e) the optical image of the
sample F4 with invasive lead contacts. The scale bar is 2 $\mu$m.
(f) Schematic band structure near a surface of our TI
samples. The crossed lines (red online) represent the TSS. $t_{a}$ represents
the range of surface band bending (or the range of carrier
accumulation) on the surface. Horizontal thick lines at $t_a$ (blue online) represent the 2DEG
formed at the surface due to the surface band bending.
Double-parabolic curves (blue online) are Rashba-split bands of the 2DEG.
Horizontal dashed line corresponds to the Fermi level.\label{ThickTemp}}
\end{figure}

The $T$ dependence of the resistivity $\rho_{xx}$ of the thick bulk
crystals of BSTS in Fig.\ \ref{ThickTemp}(a) exhibits conventional semiconducting
behavior down to $\sim 40$ K. A fit of $\rho_{xx}(T)$ to the
Arrhenius law renders the activation energy of $E_a=$ 26.1, 21.3,
31.6 and 20.7 meV for samples B1, B2, B3 and B4 (inset in Fig.\ \ref{ThickTemp}(a)
corresponds to sample B3), consistent with previous
studies. \cite{Arakane} However, the resistance is saturated for $T$
below $\sim 40$ K, which indicates the emergence of additional
conducting channels. This behavior was more pronounced in the thin
flakes. Figure\ \ref{ThickTemp}(b) shows a clear semiconductor-metal transition as
the thickness of the flakes decreases. The variation of
$\rho_{xx}(T)$ with the flake thickness can be interpreted in terms
of surface-conducting channels in the presence of a bulk insulating
gap, as illustrated in Fig.\ \ref{ThickTemp}(f). With $E_F$ inside the bulk energy
gap, the residual bulk conduction by carriers thermally activated
from an impurity band was dominant in the thick crystals (Fig.\ \ref{ThickTemp}(a)).
Thin flakes, however, with less bulk conductance, exhibited
metallic behavior. One can confirm this behavior by modelling the
simple form for total sheet conductance as follows:
\begin{equation}\label{SheetConductance}
G_\square=G_s+\sigma_b t
\end{equation}
where $G_s$ is the surface sheet conductance, $\sigma_b$ is the bulk
conductivity, and $t$ is the thickness of crystals. Here, $G_s$
includes the conduction through the 2DEG layer (see Fig.\ \ref{ThickTemp}(f)) in
the potential well formed by surface band bending, as well as the
conduction by the TSS. \cite{Bianchi,King,BeniaPRL177602} Fitting
the observed results to Eq.\ (\ref{SheetConductance}) (Fig.\ \ref{ThickTemp}(c)), $\sigma_b$ is estimated
to be $86.9$ and $26.7$ $(e^2/h)$ $\mu$m$^{-1}$ at 290 and 4.2 K,
respectively. These values are at least two orders of magnitude
smaller than the ones reported previously for Bi$_2$Se$_3$,
\cite{SteinbergNano5032} indicating that our BSTS single crystals
were highly \textquotedblleft bulk-insulating\textquotedblright.
Assuming the range of surface band bending at the surface to be
$\sim$30 nm (see Appendix\ \ref{B}) in sample F4, the
relative weight of the bulk to the surface conductance becomes
$\sigma_b t/G_s\sim 26\%$ ($6\%$) at 290 K (4.2 K).

\subsection{Angle and temperature dependence of WAL and UCF}

\begin{figure}
\includegraphics[scale=0.48]{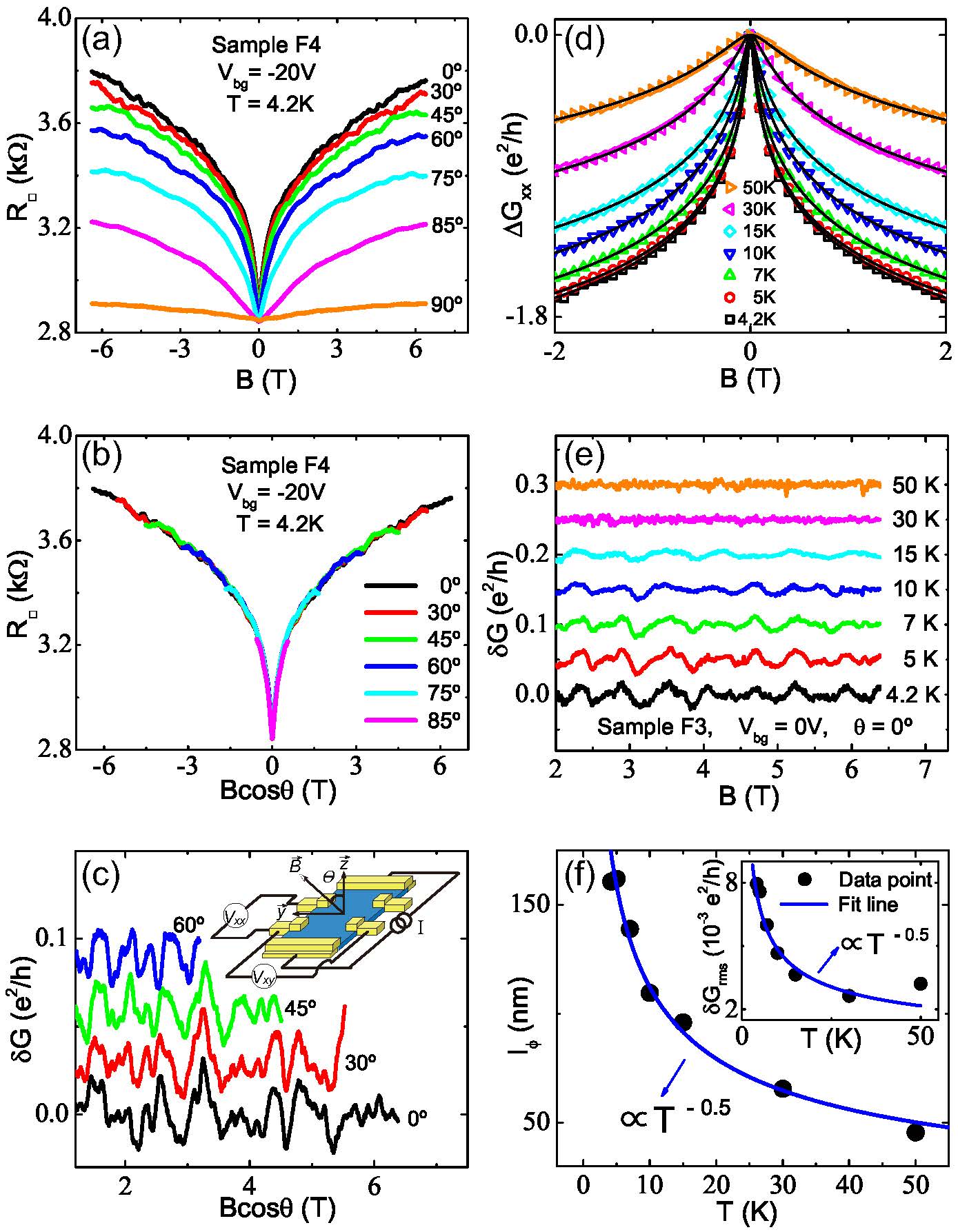}
\caption{(Color online) (a) Sheet resistance $R_\square~$vs$~B$ and
(b) $R_\square~$vs$~B\cos\theta$ for different field angles
at $V_{bg}=-20$ V for sample F4. (c) $\delta G=G(B)-G_0(B)$
vs $B\cos\theta$ extracted from the data in (b), where
$G_0(B)$ is the background of $G(B)$ (=$1/R_\square(B)$). For
clarity, each curve is shifted vertically by 0.03 $e^2/h$.
Inset: Schematic measurement configuration.
(d) $\Delta G_{xx}(B)=G_{xx}(B)-G_{xx}(B=0)~$vs$~B$ at
different $T$ for sample F3. Solid lines are best fits to Eq.\ (\ref{HLNequation}).
(e) $T$ dependence of $\delta G$ of sample F3 extracted
from the data in (d). For clarity, each curve is shifted
vertically by 0.05 $e^2/h$. (f) $T$ dependence of the
phase-relaxation length $l_\phi$ obtained from (d). Inset:
$T$ dependence of the root-mean-square of $\delta G$ extracted from
the data in (e).\label{AngleTemp}}
\end{figure}

The surface-dominant conduction at low $T$ becomes more evident in
the field-angle dependence of the magnetoresistance (MR). Figure \ref{AngleTemp}(b)
shows that all of the MR curves taken at different field angles
(Fig.\ \ref{AngleTemp}(a)), plotted as a function of the normal component of the
field (B$_\bot$), merge into a single universal curve (see Appendix \ref{C}
for the discussion on the MR feature in
in-plane fields; $\theta$=90$^\circ$). Even the positions of the UCF
peaks agree with each other when plotted as a function of B$_\bot$
(Fig.\ \ref{AngleTemp}(c)). These features strongly indicate that the MR in our
sample was almost completely dominated by surface conduction over
the entire field range of our measurements. Previously, the
cos($\theta$) angle dependence of the MR was observed only in the
low-field range of $B$ within a fraction of tesla.
\cite{ChaNanoLett1107,HePRL166805}

The 2D nature was identified more quantitatively from the $T$
dependence of the MR. Figure \ref{AngleTemp}(d) is the $T$ dependence of WAL
effects and the best fits of $\Delta G_{xx}(B)$ to Eq.\ (\ref{HLNequation}), from
which we obtained the $T$ dependence of the phase relaxation length
$l_\phi$ as shown in Fig.\ \ref{AngleTemp}(f) (more details of the WAL effect are
discussed below). Figure \ref{AngleTemp}(e) shows the $T$ dependence of $\delta
G$, with the corresponding $T$ dependence of the UCF amplitude
$\delta G_{\textrm{rms}}$ shown in the inset of Fig.\ \ref{AngleTemp}(f). In a 2D
system with a sample dimension of $L \gg l_\phi$, $l_\phi$ scales as
$T^{-0.5}$ for inelastic scattering by electron-electron
interaction, and $\delta G_{\textrm{rms}}$ is proportional to
$l_\phi$. \cite{Choi,Altshuler,UCF1,UCF2} In Fig.\ \ref{AngleTemp}(f), both
$l_\phi$ and $\delta G_{\textrm{rms}}$ scale as $T^{-0.5}$, in good
agreement with the theoretical predictions, indicating that the
dominant inelastic scattering in the surface-conducting channels of
our BSTS flakes was due to the electron-electron interaction.

\subsection{Back-gate dependence of WAL}

\begin{figure*}
\includegraphics[scale=0.8]{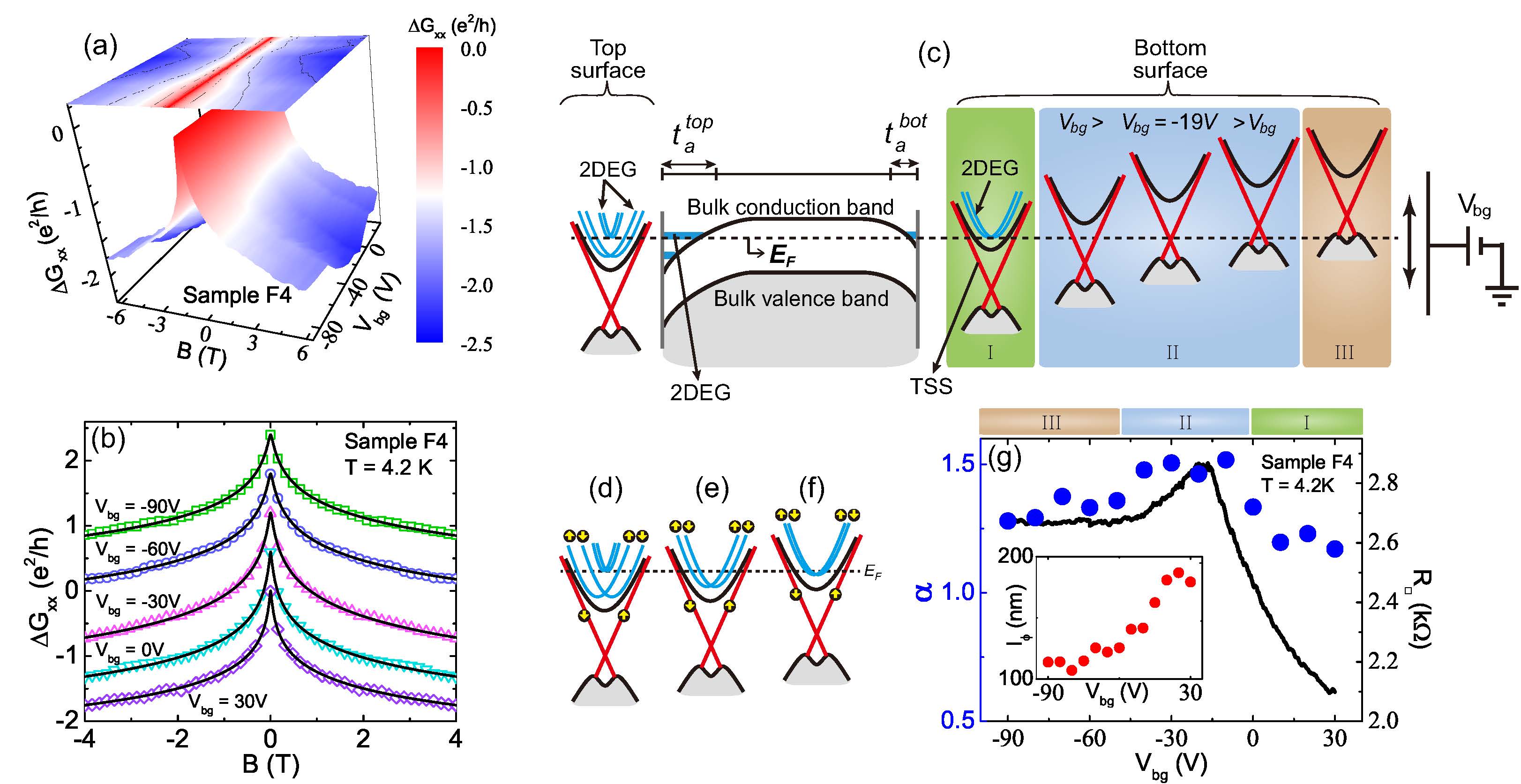}
\caption{(Color online) (a) $\Delta G_{xx}$ vs $B$ in color codes (online) as a function $V_{bg}$.
(b) $\Delta G_{xx}(B)$ curves for different
values of $V_{bg}$. Solid lines are best fits to Eq.\ (\ref{HLNequation}). For
clarity, each curve is shifted vertically by $0.6~e^2/h$.
(c) Schematic band diagram relevant to the thin flakes used
in this study. The crossed lines (red online) represent the TSS. $t^{top}_{a}$
($t^{bot}_{a}$) represents the range of the carrier accumulation on
the top (bottom) surface. Horizontal thick lines at $t^{top}_{a}$ and $t^{bot}_{a}$ (blue online) represent the 2DEG formed
at the top and bottom surface due to the surface band bending.
Double-parabolic curves (blue online) are Rashba-split bands of the 2DEG.
Horizontal dashed line depicts the Fermi level. Regions I, II, and
III represent the band structure for the corresponding regions
denoted in (g). (d, e, f) Schematic diagram of the Rashba-split strength of 2DEG for different band bending. Up and
down arrows indicate the spin texture. (g) $V_{bg}$
dependence of $\alpha$, obtained from best fits to Eq.\ (\ref{HLNequation}). Solid
curves are the $V_{bg}$ dependence of $R_\square$. Inset: $V_{bg}$
dependence of $l_\phi$, also obtained from best fits to Eq.\ (\ref{HLNequation}).\label{GateWAL}}
\end{figure*}

Up to this point, results from our BSTS consistently indicate that
the bulk conduction was negligible, and that both WAL and UCF had a
2D nature. The WAL in the TSS arose from the Berry phase $\pi$
caused by the helical spin texture. Since the Rashba-split 2DEG has
the momentum-locked spin helicity (see Figs. \ref{GateWAL}(d), (e), and (f)),
the topologically trivial 2DEG states also exhibit the WAL effect.
Applying $V_{bg}$, we confirmed that the WAL effect arose from
surface conduction, in both TSS and the topologically trivial 2DEG,
with negligible bulk conduction. According to the HLN theory, for a
2D system in the symplectic limit, \textit{i.e.}, in the limit of
strong spin-orbit coupling ($\tau_\phi\gg\tau_{so},\tau_e$;
$\tau_\phi$ is the dephasing time, $\tau_{so}$ the spin-orbit
scattering time, and $\tau_e$ the elastic scattering time) with a
negligible Zeeman term, the magnetoconductance correction is given
as follows:
\begin{equation}\label{HLNequation}
\Delta G_{xx}=\alpha\frac{e^2}{2\pi^2\hbar}
\left[\ln\left(\frac{\hbar}{4el^2_\phi
B}\right)-\psi\left(\frac{1}{2}+\frac{\hbar}{4el^2_\phi
B}\right)\right],
\end{equation}
where $\psi$ is the digamma function, $e$ is the electronic charge,
$\hbar$ is Planck's constant divided by $2\pi$, and $l_\phi$ is the
phase-relaxation length. \cite{HLN} Because the WAL effect
constitutes a prominent transport property of the TSS, the
relationship between the parameter $\alpha$ and the number of
conducting channels in the symplectic limit is essential to
differentiating the transport nature of TIs. \cite{Garate} Each 2D
conducting channel in the symplectic limit contributes 0.5 to the
value of $\alpha$. If there are two independent 2D conducting
channels in the symplectic limit, $\alpha=\alpha_1+\alpha_2$
($\alpha_i$, corresponding to the channel $i$) and $l_\phi$ is
replaced by the effective phase relaxation length (see Appendix \ref{D} for details of the WAL fitting).

We confirmed that the back-gating affected only the bottom-surface
conductance for the 85$\sim$90 nm-thick samples (F3 and F4) (see Appendix \ref{E}).
Figure \ref{GateWAL}(a) shows $\Delta G_{xx}$ vs $B$
in color codes (online) as a function $V_{bg}$. Here, the WAL effect occurs
over the entire range of $V_{bg}$ of this study with a maximum
$\Delta G_{xx}$ at $V_{bg}\sim-19$ V, the Dirac point of the TSS at
the bottom surface (corresponding to the center diagram in Region II
of Fig.\ \ref{GateWAL}(c)). Figure \ref{GateWAL}(b) shows $\Delta G_{xx}$ curves for
different values of $V_{bg}$, which agree well with Eq.\ (\ref{HLNequation}) (solid
curves) over the entire range of $B$; the corresponding values of
$\alpha$ are plotted in Fig.\ \ref{GateWAL}(g). For all $V_{bg}$, $\alpha$
exceeds unity, indicating that more than two 2D conducting channels
with the symplectic-limit behavior were involved in the surface
conduction.

In Region II of Fig.\ \ref{GateWAL}(g), the TSS in the bottom surface contributes
a value of 0.5 to $\alpha$. This leaves $\alpha \sim 1$ for the top
surface, which does not appear to be affected by $V_{bg}$. Thus, we
infer that the band bending near the top surface is like what is
shown in Fig.\ \ref{GateWAL}(c). In the top surface, in addition to the TSS, the
two Rashba-split channels in the trivial 2DEG layer also exhibit WAL
in the symplectic limit. \cite{Bianchi,King,BeniaPRL177602} However,
the magnitude of $\alpha$ is reduced from 1.5 (=0.5$\times$3) to
$\sim$1 due to inter-band scattering, where the degree of reduction
depends on the scattering strength. \cite{Garate} In Region I, $E_F$
also enters the bulk conduction band (BCB) of the bottom surface.
But, if the surface band bending is not enough to make a sufficient
Rashba splitting in the 2DEG states as in Fig.\ \ref{GateWAL}(f), the band
structure of the 2DEG would be similar to the unitary case,
\cite{HLN} where the scattering between the TSS and the
topologically trivial 2DEG states is enhanced along with weakening
of the WAL effect. \cite{LuPRB125138} This reduces the value of
$\alpha$ of the bottom surface down to $\sim 0.2 - 0.3$, while
leaving $\alpha$ unchanged at $\sim1$ for the top surface. If
$E_{F}$ is shifted deeper into the conduction band as to form a 2DEG
on the bottom surface with a large-Rashba-split bulk subband (Fig.\ \ref{GateWAL}(d)),
the WAL effect will be enhanced again, with the value of
$\alpha$ larger than 0.5 as shown in Fig.\ \ref{GateWAL}(c) for the top surface.
\cite{Garate} In Region III, a similar reduction of $\alpha$ is
expected for the bottom surface, due to the enhanced scattering
between the TSS and the bulk valence band (BVB). Thus, the variation
of $\alpha$ with $V_{bg}$ in Fig.\ \ref{GateWAL}(g) is the result of variation of
the WAL in the bottom surface state.

The WAL effects reported previously on TIs with $\alpha\sim0.5$
\cite{ChenPRL176602,KimNatPhys460,ChenPRB241304,KimPRB073109,MatsuoPRB075440}
or $\alpha\sim1$
\cite{CheckelskyPRL196801,HongNatComm757,SteinbergPRB233101,ZhangAFM2351,ChaNanoLett1107,GaoAPL}
contained a finite bulk contribution. $\alpha\sim0.5$ corresponded
to an effective single layer formed by the bulk and the two (top and
bottom) surfaces, which are strongly coupled together. Meanwhile,
$\alpha\sim1$ corresponded to an effective single layer formed by
the $n$-type bulk strongly coupled to the top surface, in
association with the $p$-type bottom surface that was decoupled from
the bulk by the formation of the depletion layer for a large
negative value of $V_{bg}$. \cite{Garate} To the best of our
knowledge, no previous reports have shown good fits to the
symplectic-limit expression of Eq.\ (\ref{HLNequation}) for fields up to several
tesla, with $\alpha$ exceeding unity.
\cite{ChenPRL176602,CheckelskyPRL196801,HongNatComm757,
KimNatPhys460,ChenPRB241304,KimPRB073109,MatsuoPRB075440,SteinbergPRB233101,ZhangAFM2351,ChaNanoLett1107,GaoAPL}
Although the good fits of our results to Eq.\ (\ref{HLNequation}) without the Zeeman
correction may be related to the recent report of small Land\'{e}
$g$ factor in TIs,
\cite{XiongPRB045314,ChengPRL076801,HanaguriPRB081305} more studies
are required to draw a definite conclusion on the issue.

\subsection{Back-gate dependence of Hall resistivity}

\begin{figure}
\includegraphics[scale=0.6]{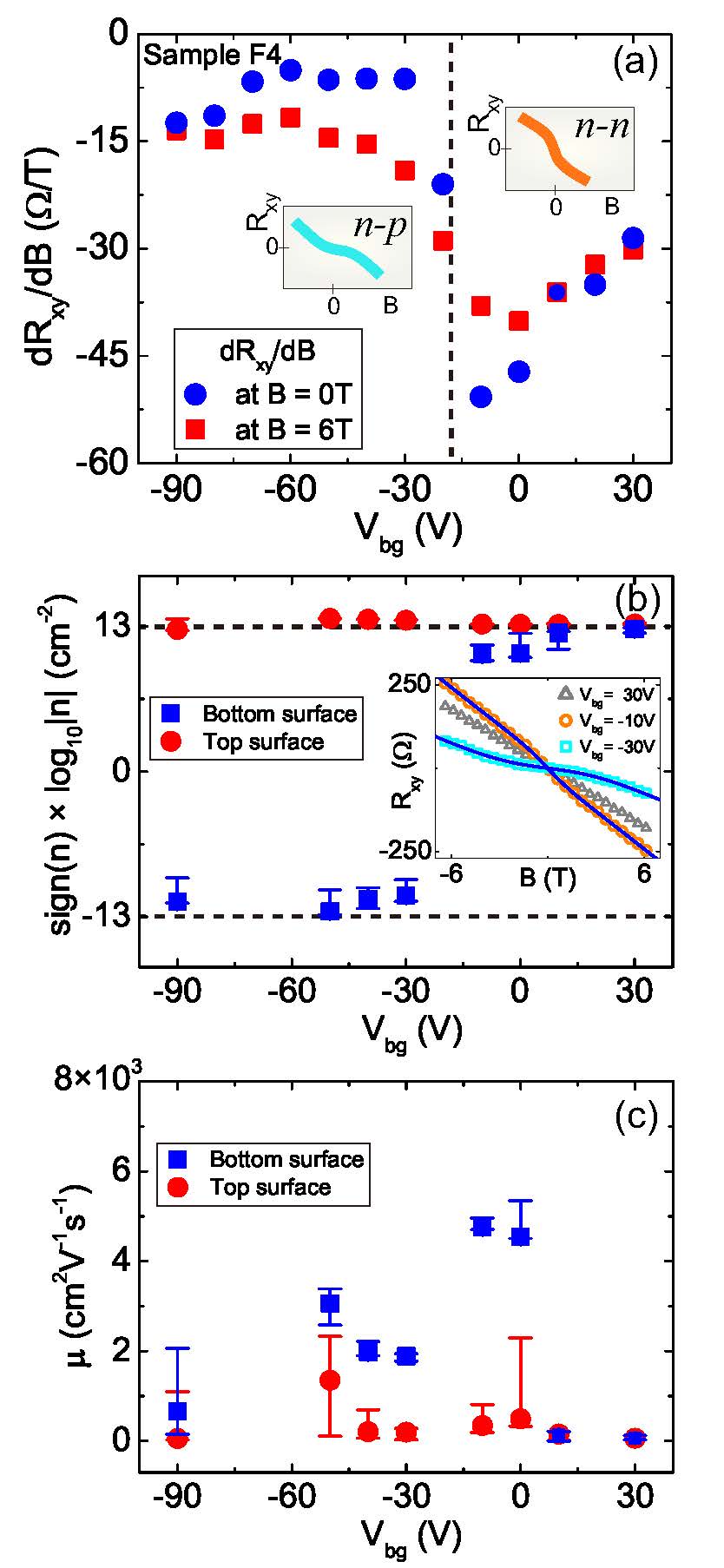}
\caption{(Color online) (a) $V_{bg}$ dependence of the slope of the
tangent to $R_{xy}(B)$ curves for Sample F4 at zero magnetic field
(circles, blue online) and $6$ T (squares, red online), respectively. The
curves in right inset (orange online) and left inset (cyan online) represent the characteristic schematic feature of nonlinear
$R_{xy}$ curve when the sample is in the
$n$-$n$ state for $V_{bg}>-19$ V and in the $n$-$p$ state for
$V_{bg}<-19$ V. The vertical dashed line represents the boundary
between the $n$-$n$ and $n$-$p$ states. (b, c) $V_{bg}$
dependencies of the carrier density and the mobility in the bottom
and top surfaces of sample F4, which are obtained by fitting the
$R_{xy}$ data to Eq.\ (\ref{twobandmodel}). The dashed lines in (b)
correspond to $\left|n\right|=10^{13}~\textrm{cm}^{-2}$. The inset
in (b) shows the representative Hall resistivity curves for
$V_{bg}=$ 30 (linear dependence), $-10$ ($n$-$n$ state, circle), and $-30$ V
($n$-$p$ state, square). The solid lines are the best fits to Eq.\ (\ref{twobandmodel}).\label{Hall}}
\end{figure}

From the thickness, field-angle, and temperature dependence of the
resistance, we conclude that in our BSTS samples the electronic
transport was dominated by the top and bottom surfaces. In this
case, the Hall resistivity can be described by a standard two-band
model as \cite{Ashcroft}
\begin{equation}\label{twobandmodel}
R_{xy}=-\left(\frac{B}{e}\right) \frac{(n_1\mu_1^2 + n_2\mu_2^2) +
B^2\mu_1^2\mu_2^2(n_1+n_2)}{(\left|n_1\right|\mu_1 +
\left|n_2\right|\mu_2)^2 + B^2\mu_1^2\mu_2^2(n_1+n_2)^2}.
\end{equation}
Here, $n_i$ and $\mu_i$ are the density and mobility of the
carriers, respectively, in the $i$-th conducting channel. The top
($i$=1) and bottom ($i$=2) surfaces constitute parallel conducting
channels, with $n_i$ being positive (negative) for $n$-type
($p$-type) carriers. \cite{Ashcroft} In sufficiently strong fields,
$R_{xy}$ converges to
$R_{xy}=R^{\textrm{strong}}_{H}B\approx-\frac{B}{e\left(n_1+n_2\right)}$.
In weak fields ($B\approx 0$ T), Eq.\ (\ref{twobandmodel}) is reduced to
\begin{align}\label{HallR}
R^{\textrm{weak}}_{H}&=R^{\textrm{strong}}_{H}\left[1+\frac{n_1n_2\left(\mu_1\mp\mu_2\right)^2}{\left(n_1\mu_1\pm
n_2\mu_2\right)^2}\right].
\end{align}
Here, the double signs are of the same order. The upper (lower) sign
corresponds to $n_1,n_2>0$ ($n_2<0<n_1$).

Figure \ref{Hall}(a) shows the results of analysis of the $V_{bg}$ dependence
of Hall resistivity from sample F4. For $V_{bg}\geq10$ V, the
difference between the square (red online) and the circle (blue online), which means nonlinearity
of $R_{xy}$, is very small, thus the $R_{xy}$ curves are almost
linear in $B$. As the $V_{bg}$ is lowered to a negative value, the
$R_{xy}$ curves starts to bend and the nonlinearity of the $R_{xy}$
increases as the $V_{bg}$ decreases (curve in right inset in Fig. \ref{Hall}(a)). In
this region (e.g. $V_{bg}=-10$ V), the slope of the tangent to
$R_{xy}$ at $B=0$ T is larger than that at $B=6$ T. However, the
feature is reversed for $V_{bg}=-$30 V (curve in left inset in Fig. \ref{Hall}(a)) and
the nonlinearity of $R_{xy}$ decreases as the $V_{bg}$ decreases.
Using Eq.\ (\ref{HallR}), it turned out that the change in the shape of the
$R_{xy}$ curve (from right inset to left inset in Fig. \ref{Hall}(a)) indicates
the ambipolar transport of Dirac fermions between the $n$-$n$ state
(top: $n$-doped, bottom: $n$-doped) and the $n$-$p$ state on TI
surfaces.

If $\mu_1\approx\mu_2$ and $n_1,n_2>0$ in Eq.\ (\ref{HallR}), then
$R^{\textrm{weak}}_{H}\approx R^{\textrm{strong}}_{H}$,
corresponding to the region of $V_{bg}\gtrsim$15 V in Fig.\ \ref{Hall}(a).
Since the $R_{xy}$ curves are almost linear in $B$, the carrier
mobility is estimated to be $\mu_1$, $\mu_2$ $\sim 140$ cm$^2$/(Vs)
using the relationship $\mu=\frac{\sigma}{ne}$, which agrees with
previous reports. \cite{TaskinPRL,XiongPhisicaE917,XiongPRB045314}
In this region (Region I in Fig.\ \ref{GateWAL}(c)) with $E_F$ in the BCB, the
mobility decreased due to the enhanced inter-band scattering.
\cite{Bianchi} As $V_{bg}$ decreased, with $E_F$ shifted to the TSS
in Region II in Fig.\ \ref{GateWAL}(c), the mobility of the bottom surface was
enhanced so that $\mu_1\neq\mu_2$. In this case, if $n_1,n_2>0$, Eq.\ (\ref{HallR})
leads to $\left|R^{\textrm{weak}}_{H}\right| >
\left|R^{\textrm{strong}}_{H}\right|$, which corresponds to the
curve for $V_{bg}=-10$ V in the inset of Fig.\ \ref{Hall}(b) (right inset in
Fig.\ \ref{Hall}(a)). Decreasing $V_{bg}$ further, $E_F$ shifted to a $p$-type
region at the bottom surface. With $n_2<0$ and
$\left|n_1\right|\gg\left|n_2\right|$, Eq.\ (\ref{HallR}) leads to
$\left|R^{\textrm{weak}}_{H}\right| <
\left|R^{\textrm{strong}}_{H}\right|$, corresponding to the curve
for $V_{bg}=-30$ V in the inset of Fig.\ \ref{Hall}(b) (left inset in Fig.\ \ref{Hall}(a)).
The change in the relative magnitude of the slopes of the
tangent to $R_{xy}(B)$, \textit{i.e.},
$\left|R^{\textrm{strong}}_{H}\right|$ and
$\left|R^{\textrm{weak}}_{H}\right|$, for $V_{bg}$ crossing $-19$ V
clearly indicates ambipolar transport of the Dirac fermions between
the $n$-$n$ and $n$-$p$ states on the TI surface. For
$V_{bg}\lesssim-50$ V (Region III in Fig.\ \ref{GateWAL}(c)), with $E_F$ in the
BVB, scattering between the TSS and the BVB was enhanced once again.
\cite{KimPRL056803} The resulting suppression of $\mu_2$, combined
with an increase of $n_2$ in the range of $V_{bg}\lesssim-50$ V
along with the relationship $\sigma_2=n_2e\mu_2$ for the bottom
surface, may explain the low sensitivity of $R_\square$ to $V_{bg}$
in Fig.\ \ref{GateWAL}(g).

Fitting the $R_{xy}$ data to Eq.\ (\ref{twobandmodel}) gives more quantitative support
for the analysis above on the $V_{bg}$ dependence of the Hall
resistivity. The inset of Fig.\ \ref{Hall}(b) shows the representative Hall
resistivity for $V_{bg}=$ 30, $-10$, and $-30$ V, where the solid
curves are the best fits to Eq.\ (\ref{twobandmodel}) with the parameter values
summarized in Figs.\ \ref{Hall}(b) and (c). $n_1$ in Fig.\ \ref{Hall}(b) is almost
constant for all values of $V_{bg}$, while $n_2$ changes its sign
between $n$ and $p$ types at $V_{bg}\sim-19$ V. This indicates
ambipolar transport for the bottom surface with varying $V_{bg}$
across the Dirac point, while the top surface remained mostly
unaffected by back-gating, consistent with earlier qualitative
analysis of $V_{bg}$ dependent Hall resistivity. This back-gating
effect on the two surfaces was also confirmed by the mobility
change. In Fig.\ \ref{Hall}(c), the best-fit values of $\mu_1$ are almost
insensitive to the variation of $V_{bg}$. However, $\mu_2$ turns out
to be significantly larger than $\mu_1$ in the region, $-$50 V$\leq
V_{bg}\leq$0 V, where $E_F$ is assumed to be in the Dirac band of
the bottom surface. The $\mu_2$ enhancement possibly stems from the
mobility increase as $E_F$ shifts into the Dirac band of the bottom
surface from the trivial 2DEG band (either conduction or valence),
where $\mu_2$ is reduced by the scattering between the TSS and the
trivial 2DEG band. It should be noted that, with the invasive
configuration of electrodes adopted in this study, the observed Hall
voltage is bound to be underestimated. However, the qualitative
$V_{bg}$ dependence of the parameters in Eq.\ (\ref{twobandmodel}) remains valid.

\section{Conclusion}
The $1/2$ Berry-phase shift in SdHO is often adopted to examine the
topological nature of surface transport. However, very strong
magnetic fields of $B\gtrsim$ 50 T with careful Landau-level
indexing, required for accurate determination of the Berry phase,
have made it difficult to clearly differentiate the conductance by
the TSS from that by the trivial 2D-conducting states. Observation
of SdHO also requires relatively high mobility with a sufficiently
long mean-free path to support the cyclotron orbital motion. In
contrast, the observation of WAL, an intrinsic 2D effect, directly
points to conduction by the TSS. Furthermore, WAL, which arises from
the coherent diffusive motion of carriers, is not limited to the
high mobility state. In this sense, the WAL effect which was used
primarily in this study can be considered to be a more essential
criterion than the SdHO for confirming the conduction by the TSS.

For flakes significantly thicker than an optimum thickness of
$\sim$80$-$90 nm, the bulk conductance cannot be neglected. On the
other hand, as the range of band bending near the top and bottom
surfaces begins to overlap for thinner flakes, independent gate
control of the surface conduction would no longer be possible. Thus,
our approach of separating the TSS by examining the transport
characteristics specific to the 2D-topological nature in the
optimal-thickness crystal flakes (in combination with back-gating)
provides a convenient means of investigating the fundamental
topological nature of the surface conduction and the quantum-device
applications associated with momentum-locked spin polarization in
the surface state of TIs.

\begin{acknowledgments}
HJL thanks V. Sacksteder for valuable discussion on the in-plane MR.
This work was supported by the National Research Foundation (NRF),
through the SRC Center for Topological Matter [Grant No.
2011-0030788 (for HJL) and 2011-0030785 (for JSK)], the GFR Center
for Advanced Soft Electronics (Grant No. 2011-0031640; for HJL), and
the Mid-Career Researcher Program (Grant No. 2012-013838; for JSK).
\end{acknowledgments}

\appendix
\section{\label{A} Possible formation of multiple parallel 2D conducting channels in TIs}
The weak anti-localization (WAL) in bulk topological insulator (TI)
single crystals and thin TI flakes with high carrier density was
reported previously.
\cite{CheckelskyPRL246601,AnalytisNatPhys960,WangNanoLett1170}
However, the magnitude of the consequent conductance correction
($\Delta G$) was larger than our results by one or two orders of
magnitude. Since the magnitude of $\Delta G$ is proportional to the
parameter $\alpha$ in Eq.\ (\ref{HLNequation}) in the main text, which corresponds to
the number of parallel conducting channels, one may suspect that
multiple two-dimensional (2D) conducting channels connected in
parallel were present for the conduction of TI in previous studies.
A recent report \cite{CaoPRL216803} supports the inference. In Ref.
[\onlinecite{CaoPRL216803}], it was concluded that the observed quantized
Hall effect and SdHO were not caused by the topologically protected
surface state (TSS) but by many topologically trivial 2D conducting
channels connected in parallel.

From the SdHO measurements, one can obtain the information on the
dimensionality and carrier density of the conducting channels. In
the SdHO analysis, the degeneracy \textquotedblleft
2\textquotedblright ~ corresponds to the bulk band or the
topologically trivial two-dimensional electron gas (2DEG) on the
surface accumulation layer, while the degeneracy \textquotedblleft
1\textquotedblright~ corresponds to the TSS. In some previous
studies, \cite{LangACSNano295,PetrushevskyPRB045131} the carrier
density was estimated from the SdHO data adopting the degeneracy
\textquoteleft 1\textquoteright~ under the assumption that the
observed SdHO arose from the TSS. The carrier density estimated in
this way was claimed to be relevant to the TSS, based on the fact
that, with $E_F$ lying in the TSS, the maximum carrier density is
expected to be $0.5\sim 0.8 \times 10^{13}$ $\textrm{cm}^{-2}$
depending on the TI materials used. However, if the SdHO had arisen
from the topologically trivial 2D conducting channels, the
degeneracy should have been \textquoteleft 2\textquoteright~ with a
doubled carrier density. In this case, however, a Dirac cone cannot
accommodate all the carrier states estimated with the degeneracy
\textquoteleft 2\textquoteright~ in Ref.
[\onlinecite{PetrushevskyPRB045131}] and
[\onlinecite{LangACSNano295}] without the bulk conduction band or
2DEG states.

In fact, the SdHO frequencies themselves obtained in Ref.\
[\onlinecite{CaoPRL216803}] and Refs.\ [\onlinecite{PetrushevskyPRB045131}, \onlinecite{LangACSNano295}]
were not much different from each other. Thus, the difference in the
carrier densities between Ref.\ [\onlinecite{CaoPRL216803}] and Refs.\
[\onlinecite{PetrushevskyPRB045131}, \onlinecite{LangACSNano295}] resulted from the
different degeneracy values adopted in the analysis. Depending on
the degeneracy value used in the SdHO analysis, one may reach very
different conclusions on the topological nature of the conducting
channels involved in the SdHO data. In this sense, observation of
the SdHO itself cannot confirm the existence of the TSS. Correctly
identifying the Berry phase in strong magnetic fields is essential
to confirming the TSS in TIs. \cite{XiongPRB045314}

\section{\label{B} Surface band bending}

The surface band bending effect is a common feature of
semiconductors. In particular, for narrow-gap semiconductors, the
transport and electronic contact properties are strongly affected by
the surface band bending. The materials which are identified as TIs
are, in general, narrow-gap semiconductors whose band gap is about
$100\sim 300~\textrm{meV}$. \cite{Zhang438} Since the energy
levels of the surface state can be shifted up to a few hundred meV,
\cite{King,BeniaPRL177602,ChenPNAS} the surface band bending has a
large influence on transport properties of TIs. But, it has not been
studied in depth to date.

The depth of the surface accumulation layer ($t_{a}$ in Fig.\ \ref{ThickTemp}(f) in
main text) depends on the distribution of the local carrier density
along the $z$-axis. \cite{AndoReview,FowlerPRL15} For samples with the relatively high
carrier density, \emph{i.e.}, if $E_F$ lies in the bulk conduction
band, $t_{a}$ was calculated to be $\sim10-25$ nm.
\cite{WangNanoLett1170,King,BeniaPRL177602,Bianchi,XiuNatNano216,KongNatNano705}
$t_{a}$ can increase further as the carrier density decreases.
\cite{AndoReview,FowlerPRL15} Since, in our sample, $E_F$ lies in
the bulk band gap with a low bulk carrier density, $t_{a}$ can be
longer than $25$nm.

In addition, in comparison with the bottom surface, the top surface
is more exposed to chemicals and e-beam irradiation through the
sample preparation processes. From the careful analysis provided in
the main text, we concluded that these processes caused the band
bending at the top surface, which was larger than that at the bottom
surface, as illustrated in Fig.\ \ref{GateWAL}(c) in the main text.

\section{\label{C} In-plane field dependence of magnetoresistance }

\begin{figure}
\includegraphics[width=1\linewidth]{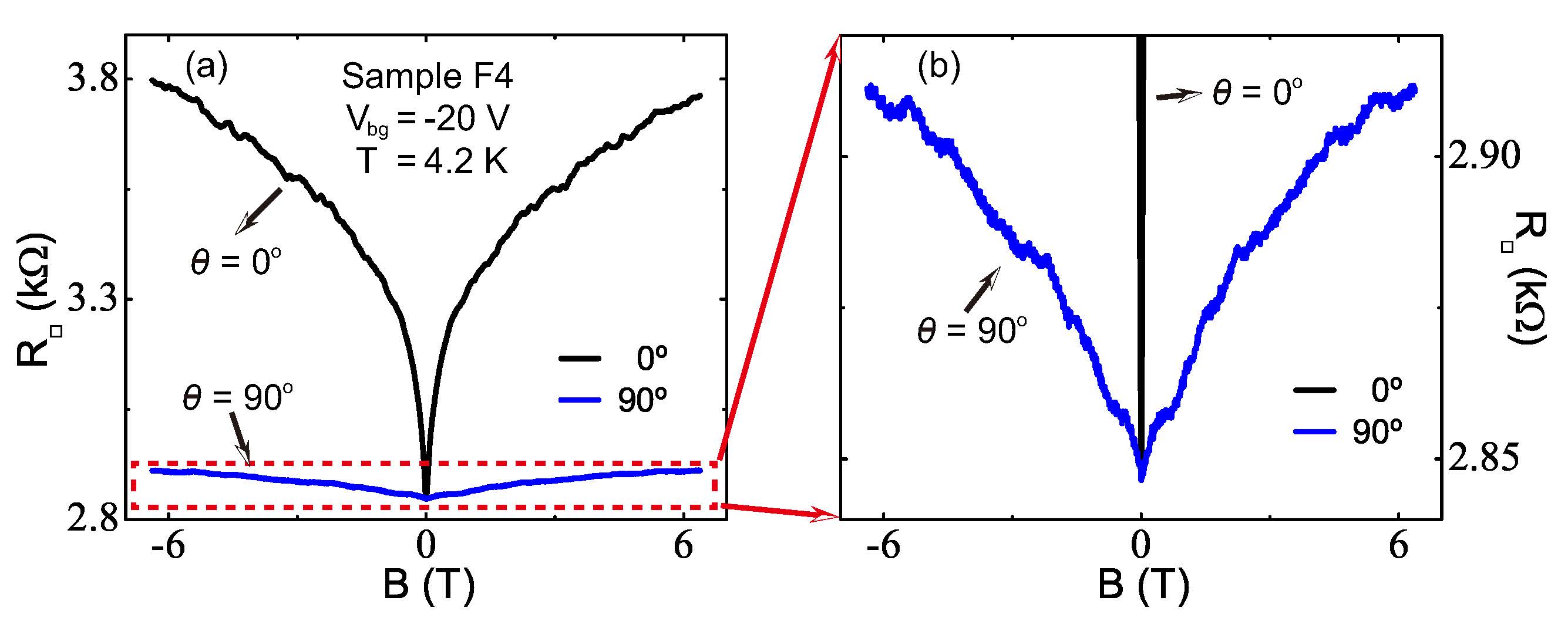}
\caption{\label{FigureS1}(Color online) (a) The MR at $V_{bg}=-20$ V
for different angles $\theta=0^\circ$ (black online) and
$\theta=90^\circ$ (blue online) of sample F4. (b) An expanded view
of the MR at $\theta =90^{\circ}$.\label{InPlaneMR}}
\end{figure}

Figure \ref{InPlaneMR} shows the MR at $\theta = 90^{\circ}$;
direction of magnetic field is in parallel with the top and bottom
surfaces of the sample and perpendicular to the current direction
(see Fig.\ \ref{AngleTemp}(a) in main text). If the conduction in our thin
flakes was only through the two surfaces (top and bottom) and only
the localization effect affected the MR, the MR should have vanished
at $\theta = 90^{\circ}$. As shown in Fig. \ref{InPlaneMR}, however,
a small but finite MR exists at $\theta = 90^{\circ}$.

The simplest inference is that the MR at $\theta = 90^{\circ}$ is a
bulk component. In Ref.\ [\onlinecite{HePRL166805}], the MR proportional to
$\sim B^2$ at $\theta = 90^{\circ}$ was observed. In the data
analysis, this component was subtracted from the MR obtained in
other field angles. Since the samples used in Ref.\ [\onlinecite{HePRL166805}]
had a large carrier density, the large weight of
the bulk conductance was reasonable with the $\sim B^2$ classical
behavior of the MR supporting that analysis.

However, in our samples, as shown in Fig.\ \ref{InPlaneMR}, we did not
find a valid argument to consider the MR at $\theta=90^\circ$ as the
three-dimensional (3D) bulk contribution. The $\sim B^2$-type
classical MR was absent at $\theta=90^\circ$. Instead, the MR
behavior was reminiscent of the WAL effect. But, there is no
consensus yet on whether the magnetoconductance (MC) correction
($\Delta G$) of bulk carriers in TIs should follow the WAL or the
weak localization (WL) behavior. \cite{LuPRB125138,Garate} Thus, it
is not clear whether the WAL-like $\Delta G(\theta=90^\circ)$ in our
data is of bulk origin.

If the MR at $\theta = 90^{\circ}$ corresponds to the 3D bulk
contribution, in order to extract $\Delta G$ of the surface
conducting channels, one has to use $\Delta G
(\theta=0^\circ)-\Delta G (\theta = 90^\circ)$ rather than $\Delta G
(\theta=0^\circ)$ as used in the main text. But the bulk origin of
$\Delta G (\theta = 90^\circ)$ is not clear. On the other hand, the
magnitude of $\Delta G$ at $\theta =90^\circ$ is sufficiently
smaller than that at $\theta = 0^\circ$ so that the discussion on
the angle dependence of MR and the $V_{bg}$ dependence of MR at
$\theta = 0^\circ$ in the main text is not affected even without
subtracting $\Delta G (\theta=90^\circ)$. Therefore, we used the raw
data for analysis of the gate dependence of WAL effects in Fig.\ \ref{GateWAL} in
main text.

It is not clear what caused this finite MR at $\theta = 90^{\circ}$
in our TI flakes. It may have arisen from the side-wall surfaces of
the thin crystal or even the in-plane MR of the surface conducting
channels. There are some theoretical prediction of in-plane
field-dependence MC correction for a 2D system, but not in the
symplectic case. \cite{Maekawa,Al'tshuler} To the best of our
knowledge, however, in-plane field-dependence MC correction of 2D
systems in the symplectic limit has not been studied yet.

\section{\label{D} Weak anti-localization analysis}

\begin{figure}
\includegraphics [scale=0.35]{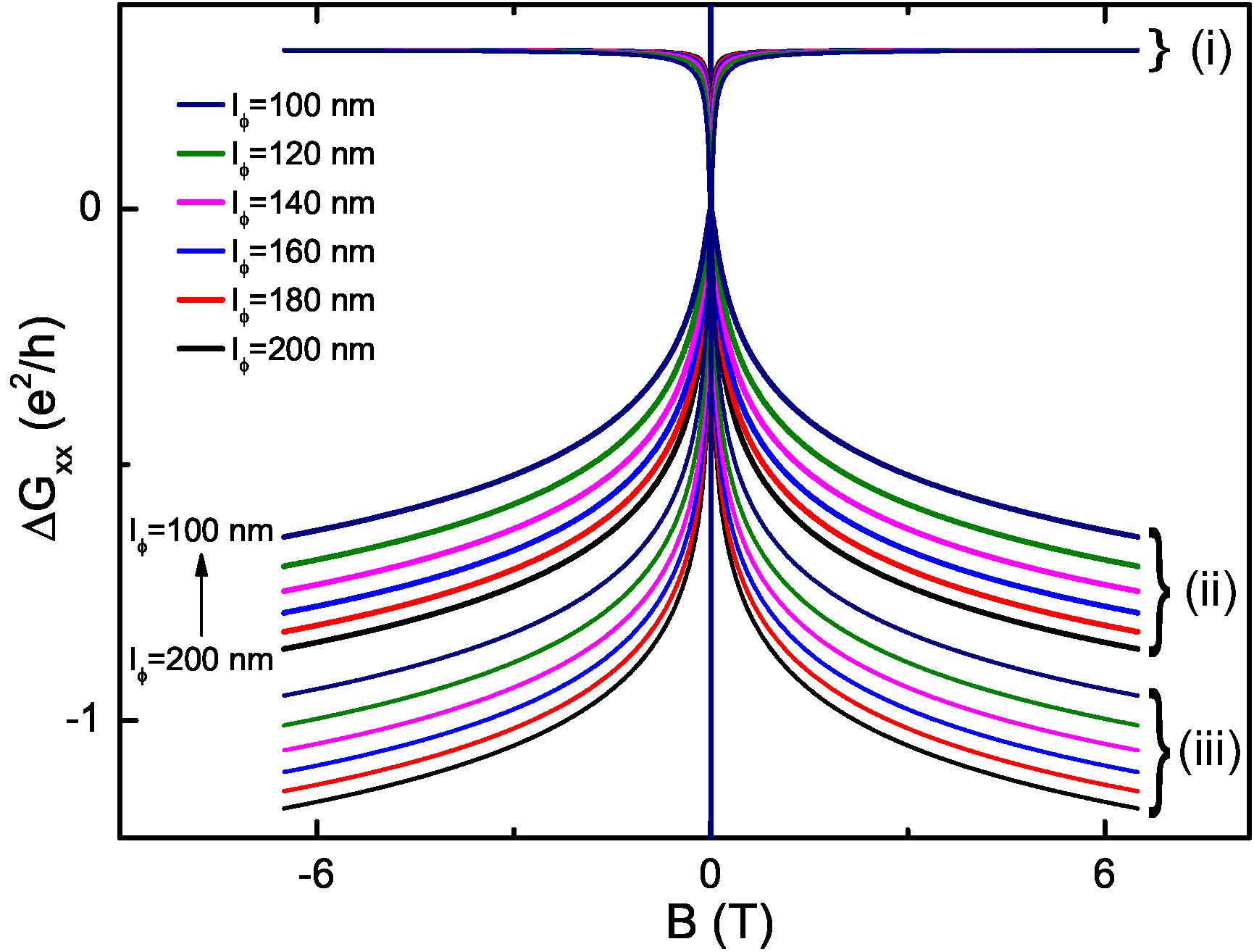}
\caption{\label{FigureS2}(Color online) Sets of graphs of (i) the digamma function
$-\frac{\alpha}{\pi}\psi\left(\frac{1}{2}+\frac{\hbar}{4el^2_{\phi}
B}\right)$, (ii) the HLN function
$\frac{\alpha}{\pi}\left[\ln\left(\frac{\hbar}{4el^2_{\phi}
B}\right)-\psi\left(\frac{1}{2}+\frac{\hbar}{4el^2_{\phi}
B}\right)\right]$, and (iii) the logarithmic function
$\frac{\alpha}{\pi}\ln\left(\frac{\hbar}{4el^2_{\phi} B}\right)$ of
Eq. (S1) for $i=1$, with $\alpha=0.5$ and $l_\phi=100$ $\textrm{nm}
\sim 200$ $\textrm{nm}$.\label{HLNfunction1}}
\end{figure}

Magnetoconductance (MC) correction of a 2D system in a symplectic
limit can be expressed as Eq.\ (\ref{HLNequation}) in the main text (HLN function).
If there are two independent 2D conducting channels, the equation is
expanded as follows:
\begin{equation}\label{TwoHLN}
\Delta G_{\textrm{xx}}=\sum_{i=1,2}\alpha_i \frac{e^2}{2\pi^2\hbar}
\left[\ln\left(\frac{\hbar}{4el^2_{\phi,i}
B}\right)-\psi\left(\frac{1}{2}+\frac{\hbar}{4el^2_{\phi,i}
B}\right)\right],
\end{equation}
where $\psi$ is the digamma function, $e$ is the electron charge,
$\alpha_i$ corresponds to the channel $i$ with the phase relaxation
length $l_{\phi,i}$. \cite{HLN} If $l_{\phi,1}=l_{\phi,2}=l_{\phi}$,
Eq. (\ref{TwoHLN}) is simplified as follows:
\begin{equation}\label{TwoHLN2}
\Delta
G_{\textrm{xx}}=\left(\alpha_1+\alpha_2\right)\frac{e^2}{2\pi^2\hbar}
\left[\ln\left(\frac{\hbar}{4el^2_{\phi}
B}\right)-\psi\left(\frac{1}{2}+\frac{\hbar}{4el^2_{\phi}
B}\right)\right].
\end{equation}
However, if $l_{\phi,1}\neq l_{\phi,2}$, the number of fitting
parameters increases up to 4 with a larger standard error. We solved
this problem by taking the following simple approximation.

In Fig.\ \ref{HLNfunction1}, the set of curves (i) represents the digamma
function part, the set (iii) corresponds to the logarithmic function
part, and the set (ii) corresponds to the sum of the two parts. Each
function is plotted with $\alpha=0.5$ and $l_\phi=100$ $\sim
 200$ $\textrm{nm}$. As displayed in Fig.\ \ref{HLNfunction1}, the digamma-function part
is almost constant except in the weak-field region for different
values of $l_\phi$. Thus, the HLN expression is mostly determined by
the logarithmic part. The digamma function causes a constant shift
of the logarithmic function and removes the logarithmic divergence
in zero field. Based on this fact, four parameters in Eq.\ (\ref{TwoHLN})
can be reduced to two parameters as follows.

Let's define $l_{\phi,i}$ is the phase relaxation length of channel
$i$ ($i=1,2$) with the corresponding coefficient $\alpha_i$ and
$l^{\textrm{eff}}_\phi$ is the effective phase relaxation length
with
$\textrm{min}\{l_{\phi,1},l_{\phi,2}\}<l^{\textrm{eff}}_\phi<\textrm{max}\{l_{\phi,1},l_{\phi,2}\}$.
Applying the approximated behavior of the digamma function leads to

\begin{align}\nonumber
\Delta
G_{\textrm{xx}}=~&\alpha_1\frac{e^2}{2\pi^2\hbar}\left[\ln\left(\frac{\hbar}{4el^2_{\phi,1}
B}\right)-\psi\left(\frac{1}{2}+\frac{\hbar}{4el^2_{\phi,1}
B}\right)\right]\\\nonumber
&+\alpha_2\frac{e^2}{2\pi^2\hbar}\left[\ln\left(\frac{\hbar}{4el^2_{\phi,2}
B}\right)-\psi\left(\frac{1}{2}+\frac{\hbar}{4el^2_{\phi,2}
B}\right)\right]\\\nonumber \approx ~
&\alpha_1\frac{e^2}{2\pi^2\hbar}\left[\ln\left(\frac{\hbar}{4el^2_{\phi,1}
B}\right)-\psi\left(\frac{1}{2}+\frac{\hbar}{4e(l^{\textrm{eff}}_\phi)^2B}\right)\right]\\\nonumber
&+\alpha_2\frac{e^2}{2\pi^2\hbar}\left[\ln\left(\frac{\hbar}{4el^2_{\phi,2}
B}\right)-\psi\left(\frac{1}{2}+\frac{\hbar}{4e(l^{\textrm{eff}}_\phi)^2B}\right)\right]\\\nonumber
=~&-\left(\alpha_1+\alpha_2\right)\frac{e^2}{2\pi^2\hbar}\psi\left(\frac{1}{2}+\frac{\hbar}{4e(l^{\textrm{eff}}_\phi)^2B}\right)\\\label{approx}
&+\frac{e^2}{2\pi^2\hbar}\left[\alpha_1\ln\left(\frac{\hbar}{4el^2_{\phi,1}
B}\right)+\alpha_2\ln\left(\frac{\hbar}{4el^2_{\phi,2}
B}\right)\right].
\end{align}

\begin{figure}
\includegraphics [scale=0.3]{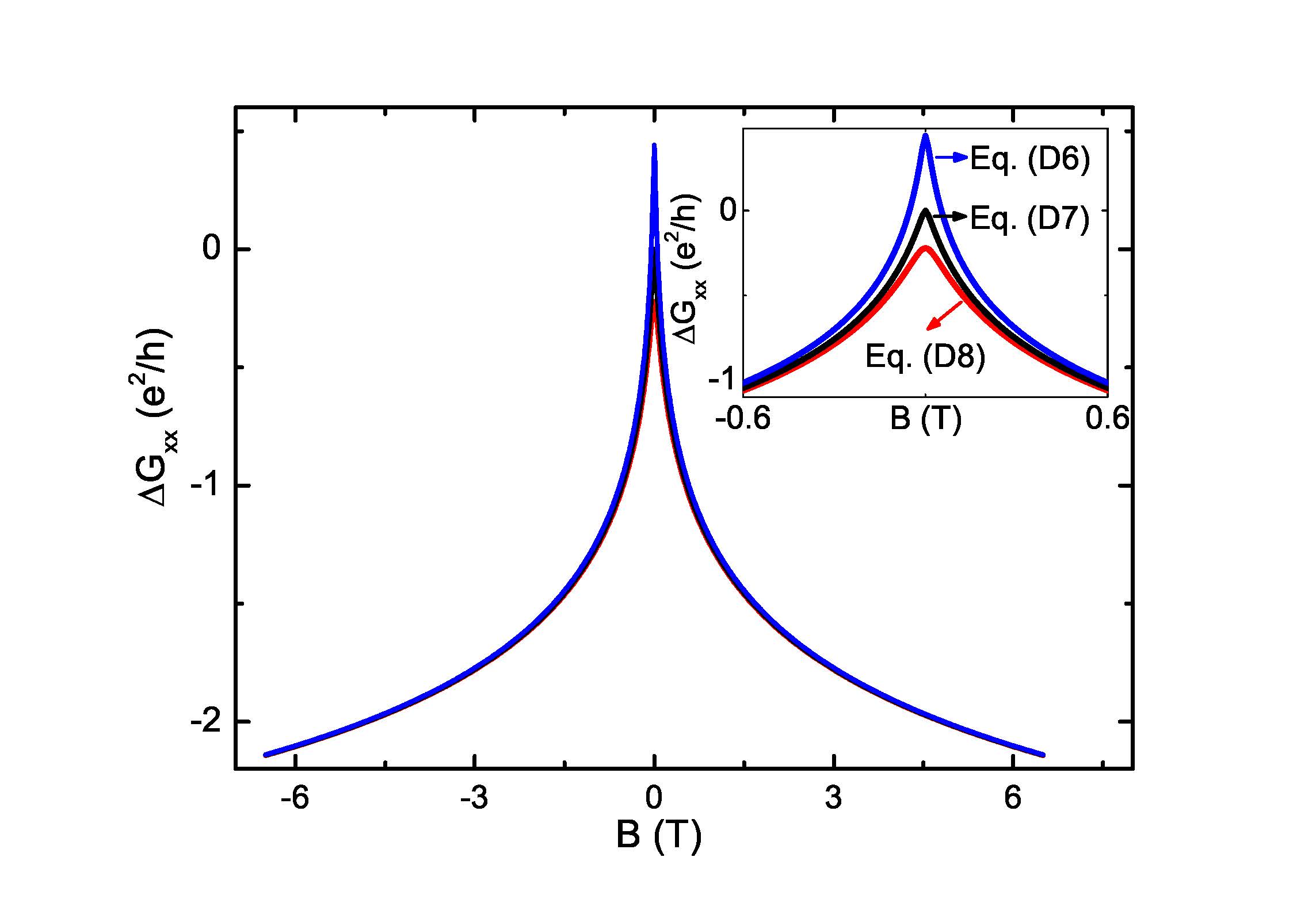}
\caption{\label{FigureS3}(Color online) Three curves (blue, black and red online) correspond to
graphs of Eqs. (D6), (D7), and (D8)  for $\alpha_1=1$,
$\alpha_2=0.5$, $l_{\phi,1}=100~\textrm{nm}$ and
$l_{\phi,2}=200~\textrm{nm}$. Inset shows the expanded view in the
low-field range.\label{HLNfunction2}}
\end{figure}

The logarithmic part in Eq.\ (\ref{approx}) becomes
\begin{align}\nonumber
&\alpha_1\ln\left(\frac{\hbar}{4el^2_{\phi,1}
B}\right)+\alpha_2\ln\left(\frac{\hbar}{4el^2_{\phi,2}
B}\right)\\\nonumber
=&\left(\alpha_1+\alpha_2\right)\frac{\alpha_1}{\left(\alpha_1+\alpha_2\right)}\ln\left(\frac{\hbar}{4el^2_{\phi,1}
B}\right)\\\nonumber
&+\left(\alpha_1+\alpha_2\right)\frac{\alpha_2}{\left(\alpha_1+\alpha_2\right)}\ln\left(\frac{\hbar}{4el^2_{\phi,2}
B}\right)\\\nonumber
=&\left(\alpha_1+\alpha_2\right)\left[\ln\left(\frac{\hbar}{4el^2_{\phi,1}B}\right)^{\frac{\alpha_1}{\left(\alpha_1+\alpha_2\right)}}+\ln\left(\frac{\hbar}{4el^2_{\phi,2}B}\right)^{\frac{\alpha_2}{\left(\alpha_1+\alpha_2\right)}}\right]\\\label{eff}
=&\left(\alpha_1+\alpha_2\right)\ln\left(\frac{\hbar}{4e(l^{\textrm{eff}}_\phi)^2B}\right)
\end{align}
where $l^{\textrm{eff}}_\phi \equiv
l^{\frac{\alpha_1}{\left(\alpha_1+\alpha_2\right)}}_{\phi,1}
l^{\frac{\alpha_2}{\left(\alpha_1+\alpha_2\right)}}_{\phi,2}$.
Therefore, with Eq.\ (\ref{approx}), the Eq.\ (\ref{TwoHLN}) can be simplified as
\begin{align}\label{WALfunction}
\Delta G_{\textrm{xx}}=~&\alpha\frac{e^2}{2\pi^2\hbar}\left[\ln\left(\frac{\hbar}{4e(l^{\textrm{eff}}_\phi)^2B}\right)-\psi\left(\frac{1}{2}+\frac{\hbar}{4e(l^{\textrm{eff}}_\phi)^2B}\right)\right]
\end{align}
with $\alpha \equiv \alpha_1 + \alpha_2$. Figure \ref{HLNfunction2} shows the
validity of this approximation.

In Fig.\ \ref{HLNfunction2}, the three curves (blue, black and red online) correspond
to the followings
\begin{align}
\alpha_1[\ln(l_{\phi,1})-\psi(l_{\phi,2})]&+\alpha_2[\ln(l_{\phi,2})-\psi(l_{\phi,2})]\\
\alpha_1[\ln(l_{\phi,1})-\psi(l_{\phi,1})]&+\alpha_2[\ln(l_{\phi,2})-\psi(l_{\phi,2})]\\
\alpha_1[\ln(l_{\phi,1})-\psi(l_{\phi,1})]&+\alpha_2[\ln(l_{\phi,2})-\psi(l_{\phi,1})],
\end{align}
respectively. Here, $\psi\left(l_{\phi,i}\right) \equiv \psi\left(\frac{1}{2}+\frac{\hbar}{4el^2_{\phi,i}B}\right)$ and $\ln\left(l_{\phi,i}\right) \equiv \ln\left(\frac{\hbar}{4el^2_{\phi,i}B}\right)$.
As displayed in Fig.\ \ref{HLNfunction2}, the deviation caused by different $l_\phi$
in digamma function can be recognized only in low fields.
Furthermore, since $l^{\textrm{eff}}_\phi$ has a value between
$l_{\phi,1}$ and $l_{\phi,2}$, the deviation may be smaller than
differences displayed in Fig.\ \ref{HLNfunction2}. Therefore, even with 4 parameters
in different two channels, we can apply the one-channel HLN function
with two parameters and the determined $\alpha$ and $l_\phi$ can be
understood as $\alpha = \alpha_1 + \alpha_2$ and $l_\phi =
l^{\textrm{eff}}_\phi$ as Eq.\ (\ref{WALfunction}).

\section{\label{E} $V_{bg}$ independence of the top-surface conductance}
\begin{figure}[t]
\includegraphics[scale=0.3]{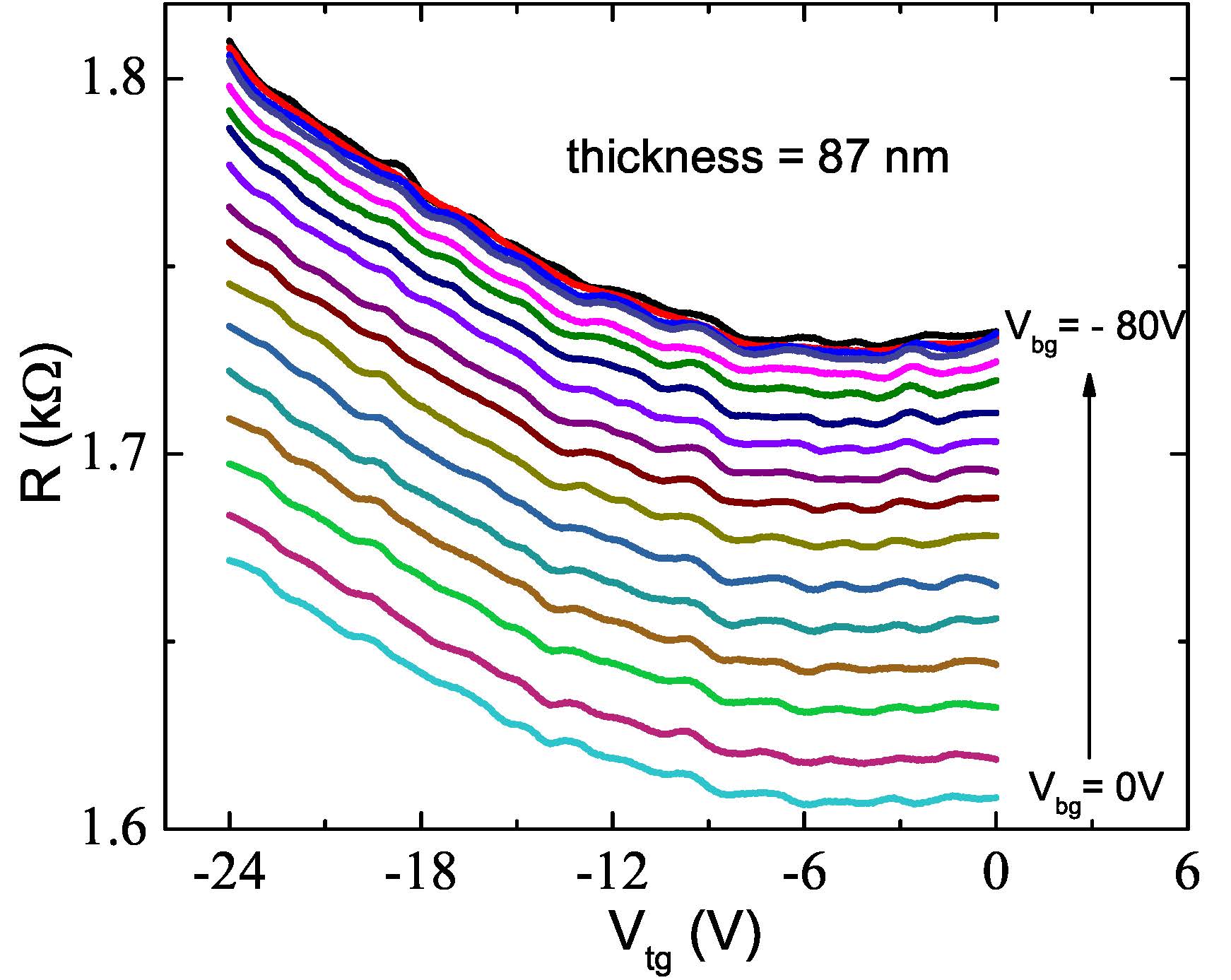}
\caption{\label{FigureS4}(Color online) $V_{bg}$ and $V_{tg}$ dependencies of the
resistance for the 87 nm-thick BSTS flake. Between two adjacent
curves $V_{bg}$ is varied by 5 V.\label{top}}
\end{figure}

Figure \ref{top} shows the resistance variation of an
87-nm-thick BSTS flake (thickness of this flake is almost identical
to that of the samples F3 and F4) as functions of back-gate
($V_{bg}$) and top-gate ($V_{tg}$) voltages. This sample is not
referred to in the main text. Except for the parallel shift in the
resistance, the $V_{tg}$ dependence of the resistance curves in Fig.\ \ref{top}
remains unaltered with varying $V_{bg}$. Even the
positions of the resistance spikes arising from the UCF effect do
not change for different values of $V_{bg}$. This feature indicates
that the top-surface (bottom-surface) conductance is almost
completely independent of $V_{bg}$ ($V_{tg}$). Since this flake and
the samples F3 and F4 are of almost identical thickness we expect
that the top-surface conductance of the two samples was independent
of $V_{bg}$, the fact of which is utilized in our analysis in the
main text.

\end{document}